\documentclass{ws-p8-50x6-00}

\def\beq{\begin{equation}}
\def\eeq{\end{equation}}
\def\beqa{\begin{eqnarray}}
\def\eeqa{\end{eqnarray}}

\def\lsim{\mathrel{\raise.3ex\hbox{$<$\kern-.75em\lower1ex\hbox{$\sim$}}} }
\def\gsim{\mathrel{\raise.3ex\hbox{$>$\kern-.75em\lower1ex\hbox{$\sim$}}} }

\renewcommand{\rightnote}{\it Talk at SUSY'01}

\begin{document}
\thispagestyle{empty}


\begin{flushright}
NCU-HEP-k001\\
Aug 2001
\end{flushright}

\vspace*{.5in}
\begin{center}
{\Large \bf \boldmath  Neutrino Masses and $\mu \to  e \, \gamma$ in the Generic Supersymmetric Standard Model$^\star$}
\vspace*{.5in}

{\bf  Otto C.W. Kong}\\[.08in]
{\it Department of Physics, National Central University, \\Chung-li, TAIWAN 32054}\\[.05in]
{\it Institute of Physics, Academia Sinica, Nankang, Taipei, TAIWAN 11529}
\vspace*{.5in}

\vspace*{.8in}
{Abstract}\\ 
\end{center}

We summarized our report on neutrino masses and $\mu \to e \, \gamma$ 
in the generic supersymmetric standard model, emphasizing on the much
overlooked scalar masses contributions from R-parity violation. 

\vfill
\noindent --------------- \\
$^\star$ Talk presented  at SUSY'01 conference (Jun 11 - 17), Dubna, Russia 
\\
 --- submission for the proceedings.  
 
\clearpage
\addtocounter{page}{-1}

\title{\protect Neutrino Masses and $\mu \to e \, \gamma$ in the Generic Supersymmetric Standard Model}

\author{Otto C. W. Kong}

\address{Institute of Physics, Academia Sinica, Nankang, Taipei, TAIWAN 11529\\
Department of Physics, National Central University, Chung-li, TAIWAN 32054$\!$
\footnote{Permanent address since August 1, 2001.}
\\
E-mail: otto@phy.ncu.edu.tw
}

\maketitle

\abstracts{We summarzied our report on neutrino masses and $\mu \to e \, \gamma$ 
in the generic supersymmetric standard model, emphasizing on the much
overlooked scalar masses contributions from R-parity violation. }

\section{On Lepton Number Violation}
The overall lepton number together with its constituents of the three lepton
flavor numbers ($L=L_e+L_\mu+L_\tau$) are accidental global symmetries of
the Standard Model (SM). Experimental data does give some support to the
idea that they are pretty good symmetries. A list of illustrative bounds
are given by 
$Br(\mu \to e \, \gamma) < 1.2 \times 10^{-11}$,
$Br(\tau \to e \, \gamma) <  2.7 \times 10^{-6}$, 
$Br(\tau \to \mu \, \gamma) < 1.1 \times 10^{-6}$, 
$Br(\mu \to 3e) <  1.2 \times 10^{-12}$, and 
$Br(\tau \to \ell_{\scriptscriptstyle 1} \, \ell_{\scriptscriptstyle 2} \,
\ell_{\scriptscriptstyle 3}) \lsim 2 \times 10^{-6}$. 
These show the stringent constraints on the possible magnitude of
lepton flavor violation (LFV). However, the strong hints of 
neutrino masses and mixings from the various experiments say otherwise.
The SM itself does not allow neutrino mass. And apart from the highly 
unlikely scenario of pure Dirac masses for the neutrinos, an extension of
the SM giving rise to neutrino masses and mixings has to violate the
lepton numbers. The present talk focuses on the minimal supersymmetric
framework to describe such violations.

\section{The Generic Supersymmetric Standard Model}
A theory built with the minimal superfield spectrum incorporating the 
SM particles, the admissible renormalizable interactions dictated by the 
SM (gauge) symmetries together with the idea that SUSY is softly broken is 
what should be called the generic supersymmetric standard model (GSSM). 
The popular minimal supersymmetric standard model (MSSM) differs from the 
generic version in having a discrete symmetry, called R parity, imposed by 
hand to enforce baryon and lepton number conservation.  The GSSM contains all 
kinds of (so-called) R-parity violating (RPV) parameters, including the 
superpotential, as well as soft SUSY breaking parameters. In order not 
to miss any plausible RPV phenomenological features, it is important that all of 
the RPV parameters be taken into consideration without {\it a priori} bias. 
We expect the lepton numbers to be violated, but are otherwise ignorant about
which admissible Lagrangian terms are mainly responsible to the resulted
phenomenology. We do, however, expect some sort of symmetry principle to guard 
against the very dangerous proton decay problem.

The renormalizable superpotential for the GSSM can be written as
\small\beqa
W \!\! &=& \!\varepsilon_{ab}\Big[ \mu_{\alpha}  \hat{H}_u^a \hat{L}_{\alpha}^b 
+ h_{ik}^u \hat{Q}_i^a   \hat{H}_{u}^b \hat{U}_k^{\scriptscriptstyle C}
+ \lambda_{\alpha jk}^{\!\prime}  \hat{L}_{\alpha}^a \hat{Q}_j^b
\hat{D}_k^{\scriptscriptstyle C} 
\nonumber \\
&+&
\frac{1}{2}\, \lambda_{\alpha \beta k}  \hat{L}_{\alpha}^a  
 \hat{L}_{\beta}^b \hat{E}_k^{\scriptscriptstyle C} \Big] + 
\frac{1}{2}\, \lambda_{ijk}^{\!\prime\prime}  
\hat{U}_i^{\scriptscriptstyle C} \hat{D}_j^{\scriptscriptstyle C}  
\hat{D}_k^{\scriptscriptstyle C}   \;.
\eeqa\normalsize
We use here the single-VEV parametrization\cite{ru,as8} (SVP), in which 
flavor bases are chosen such that : 
1/ among the $\hat{L}_\alpha$'s, only  $\hat{L}_0$, bears a VEV,
{\it i.e.} {\small $\langle \hat{L}_i \rangle \equiv 0$};
2/  {\small $h^{e}_{jk} (\equiv \lambda_{0jk}) 
=\frac{\sqrt{2}}{v_{\scriptscriptstyle 0}} \,{\rm diag}
\{m_{\scriptscriptstyle 1},
m_{\scriptscriptstyle 2},m_{\scriptscriptstyle 3}\}$};
3/ {\small $h^{d}_{jk} (\equiv \lambda^{\!\prime}_{0jk} =-\lambda_{j0k}) 
= \frac{\sqrt{2}}{v_{\scriptscriptstyle 0}}{\rm diag}\{m_d,m_s,m_b\}$}; 
4/ {\small $h^{u}_{ik}=\frac{\sqrt{2}}{v_{\scriptscriptstyle u}}
V_{\mbox{\tiny CKM}}^{\!\scriptscriptstyle T} \,{\rm diag}\{m_u,m_c,m_t\}$}, where 
${v_{\scriptscriptstyle 0}} \equiv  \sqrt{2}\,\langle \hat{L}_0 \rangle$
and ${v_{\scriptscriptstyle u} } \equiv \sqrt{2}\,
\langle \hat{H}_{u} \rangle$. The big advantage of the SVP is that it gives 
the complete tree-level mass matrices of all the states (scalars and fermions) the simplest structure.\cite{as5,as8} Readers are referred to my other talk at
Dubna\cite{002} and references therein for details of the model formulation.

\section{Neutrino Masses in GSSM}
Neutrino masses and oscillations is no doubt a central aspect of any
RPV model. In our opinion, it is particularly important to study
the various RPV contributions in a framework that takes no assumption
on the other parameters. Our formulation provides such a framework.
Earlier works alone the line include Refs.\cite{ok,as1,as5,AL}. We would
like to emphasize that the best strategy to study neutrino masses in GSSM
would be to admit our ignorance and listed all sources of neutrino masses
from different combinations of lepton number violating parameters.\cite{as9} 

\begin{table}[h] \begin{center}
\caption{\small \label{table2}
Illustrative bounds on  combinations 
of  RPV parameters
from $\mu\to e\,\gamma$.} 
\begin{tabular}{|lr|}\hline 
\ \ 
$\frac{|{\mu_{\scriptscriptstyle 3}^*}\,{\lambda_{\scriptscriptstyle 321}}|}
{|\mu_{\scriptscriptstyle 0}|}\;, \;\;\;
\frac{|{\mu_{\scriptscriptstyle 1}^*}\,{\lambda_{\scriptscriptstyle 121}}|}
{|\mu_{\scriptscriptstyle 0}|}\;, \;\;\;
\frac{|{\mu_{\scriptscriptstyle 3}}\,{\lambda_{\scriptscriptstyle 312}^*}|}
{|\mu_{\scriptscriptstyle 0}|}\;, \;\;\; 
\mbox{or} \;\;\;
\frac{|{\mu_{\scriptscriptstyle 2}}\,{\lambda_{\scriptscriptstyle 212}^*}|}
{|\mu_{\scriptscriptstyle 0}|}\; \;\;\;$ & 
$< 1.5 \times 10^{-7}$ \ \ 
\\ \ \
$\frac{|\mu_{\scriptscriptstyle 1}^*\, \mu_{\scriptscriptstyle 2}|}
{|\mu_{\scriptscriptstyle 0}|^2}$	&	$ < 0.53 \times 10 ^{-4}$ \ \
\\ \ \
$|\lambda_{\scriptscriptstyle 321} \lambda^*_{\scriptscriptstyle 131}|\;, \;\;\;
|\lambda_{\scriptscriptstyle 322} \lambda^*_{\scriptscriptstyle 132}|\;, \;\;\; 
\mbox{or} \;\;\;
|\lambda_{\scriptscriptstyle 323} \lambda^*_{\scriptscriptstyle 133}|$      
& $<2.2 \times 10^{-4}$ \ \ 
\\ 
\ \
$|\lambda^*_{\scriptscriptstyle 132} \lambda_{\scriptscriptstyle 131}|\;, \;\;\;
|\lambda^*_{\scriptscriptstyle 122} \lambda_{\scriptscriptstyle 121}|\;, \;\;\; 
\mbox{or} \;\;\;
|\lambda^*_{\scriptscriptstyle 232} \lambda_{\scriptscriptstyle 231}|$      
& $<1.1 \times 10^{-4}$ \ \ 
\\ 
\ \  
$\frac{|B_{3}^*\,\lambda_{\scriptscriptstyle 321}|}{|\mu_{\scriptscriptstyle 0}|^2}
\;, \;\;\;
\frac{|B_{1}^*\,\lambda_{\scriptscriptstyle 121}|}{|\mu_{\scriptscriptstyle 0}|^2}
\;, \;\;\;
\frac{|B_{3}\,\lambda_{\scriptscriptstyle 312}^*|}{|\mu_{\scriptscriptstyle 0}|^2}
\;, \;\;\; \mbox{or} \;\;\;
\frac{|B_{2}\,\lambda_{\scriptscriptstyle 211}^*|}{|\mu_{\scriptscriptstyle 0}|^2}$
&  $<2.0\times 10^{-3}$ \ \ 
\\ 
\ \ 
$\frac{|B_1^* \, \mu_{\scriptscriptstyle 2}|}{|\mu_{\scriptscriptstyle 0}|^3}$
 & $< 1.1\times 10^{-5}$  \ \ 
\\ \hline
\end{tabular}\end{center}
\end{table}

\section{$\mu \to e \, \gamma$ in GSSM}
We gave explicitly the complete tree-level scalar masses of the model in 
a recent paper.\cite{as5} Many of such RPV scalar mass contributions and
their phenomenological implications have been overlooked. We are starting
to explore into the domain.\cite{as4,as6,as7} 

A brief summary of resulted constraints from our extensive analytical and 
numerical study on $\mu \to e \, \gamma$,\cite{as7,cch2} are shown in Table~1.
In terms of LFV, the parameter combinations involved obviously have
$\Delta L_e =1$ and $\Delta L_\mu =-1$. The combinations
$\mu^*_k\,\lambda_{k{\scriptscriptstyle 21}}$, $\mu^*_k\,\lambda_{k{\scriptscriptstyle 12}}$, and 
$\mu^*_{\scriptscriptstyle 2}\, \mu_{\scriptscriptstyle 1}$ are directly
reflecting the corresponding slepton mass mixing contributions. The bounds
on a $\mu_k$ and $\lambda_{kij}$ combination are stringent even in comparison
to the sub-eV neutrino mass bounds. Exploring the correlation of the two would 
be particularly interesting.

\end{document}